\documentclass{aa}  

\usepackage{subfigure}
\usepackage{graphicx}
\usepackage{txfonts}
\usepackage{tipa}
\usepackage{caption}
\usepackage{supertabular}
\usepackage{nameref}
\usepackage{pgffor}
\usepackage{xcolor}

\newcommand{\gray}[0]{$\gamma$-ray}

\makeatletter
\renewcommand*\aa@pageof{, page \thepage{} of \pageref*{LastPage}}
\makeatother

\begin{document}

   \title{Characterization and classification of $\gamma$-ray bursts from blazars}

   \author{Matteo Cerruti
          \inst{1}
          }

   \institute{Université Paris Cité, CNRS, Astroparticule et Cosmologie, F-75013 Paris, France \\ 
              \email{cerruti@apc.in2p3.fr}
             }

   \date{Received XX; accepted XX}

% \abstract{}{}{}{}{}
% 5 {} token are mandatory
 
  \abstract
    {Blazars are the most common sources of $\gamma$-ray photons in the extragalactic sky. Their $\gamma$-ray light curves are characterized by bright flaring episodes, similarly to what is observed at longer wavelengths. These gamma-ray bursts from blazars (GRBBLs) have been extensively studied individually, but never in terms of a population.   }
   {The goal of this work is to provide a global characterization of GRBBLs, to investigate the parameter space of the population, and ultimately to classify GRBBLs. Their global properties could give insights into the physical mechanisms responsible for the $\gamma$-ray radiation and into the origin of the observed variability.}
   {I analyzed a sample of publicly available \textit{Fermi}-LAT light curves, utilizing only blazars with certain redshift measurements. The redshift-corrected light curves were then automatically scanned to identify GRBBLs. A simple flare profile, with an exponential rise and decay, was then fit to all events. The fit parameters, together with the information on spectral variability during the events, and the global properties from the LAT catalog, were then used as inputs for unsupervised machine learning classification.}
   {The analysis shows that the GRBBL population is remarkably homogeneous. The classifier splits the population into achromatic (the large majority) and chromatic (the outliers) GRBBLs, but the transition between the two classes is smooth, with significant overlap. When the information on the spectral variability is removed, there is evidence for a classification into two classes, mainly driven by the peak luminosities. As a by-product of this study, I identify a correlation between the timescales of the GRBBLs and their peak luminosity.}
   {}


   \maketitle
%
%________________________________________________________________

\section{Introduction}

Blazars are a peculiar class of active galactic nuclei (AGNs) characterized by unusual observational properties: a spectral energy distribution dominated by a nonthermal continuum that spans the whole electromagnetic spectrum, from the radio up to \gray s; a high degree of polarization in radio, optical, and X-rays; and an extreme variability that can be as fast as minutes. Within the AGN unified model, blazars are understood as radio-loud AGNs (i.e., showing a pair of relativistic jets launched by the central supermassive black-hole) whose jet points toward Earth. The observers are thus seeing primarily the emission from the jet of plasma, Doppler-boosted in their reference frame. This relativistic boost makes blazars particularly interesting: they are much brighter than their parent population of off-axis AGNs, and can be detected at high redshifts and high energies. Indeed, the \gray\ sky is dominated by blazars, which represent around 80$\%$ of all extragalactic sources in the giga-electronvolt band \citep[or, high-energy \gray s, with $100$ MeV $\leq E \leq 100$ GeV;][]{4LAC}  and of all extragalactic sources in the tera-electronvolt band  \citep[or, very-high-energy \gray s, with $100$ GeV $\leq E \leq 100$ TeV;][]{tevcat}.\\

The LAT instrument on board the \textit{Fermi} satellite \citep{Atwood09} has revolutionized our understanding of blazars thanks to its full-sky survey capabilities. Since its successful launch in 2008, we now have access to uninterrupted light curves of thousands of sources in giga-electronvolt \gray s. Blazar studies with LAT data are thus an active research topic in high-energy astrophysics. Similarly to what observed at longer wavelengths, \gray\ blazar light curves show bright flares, when the flux of the source increases by several magnitudes. These flares have been extensively studied, in particular in the context of their multiwavelength behavior (and more recently also in a multi-messenger context, using neutrino telescopes), but these kinds of works almost always focus on single sources, or at best on a select number of objects that share common properties. Systematic studies of blazar light curves have been performed, but focusing on variability properties in a larger sense, for example by investigating the power spectral densities \citep[see e.g.][]{Tarn20} or searching for periodicities \citep[see e.g.][]{Ren23QPO}. A systematic study of \gray\ blazar flares has not yet been performed. Worse still, there is currently no commonly accepted definition of what a blazar flare is. The goal of this work is to partially answer questions such as what defines a gamma-ray blazar flare, how long it typically lasts for, whether there are similarities among gamma-ray blazar flares, and whether there are well-defined classes of flares. Surprisingly, the high-energy astrophysical community -- so keen to use acronyms to identify its sources of interest (GRBs and FRBs, TDEs and AGNs, SNs and SNRs, MSPs and PWNs) -- does not have a name for a gamma-ray blazar flare and is forced to use periphrases to express the subject of its study. To ease readability, I introduce in the following the acronym GRBBL\footnote{It might not be perceived as a huge improvement, and unpronounceable, but it can be read either by spelling out the letters as we do for GRB \textipa{[""\t{dZ}i:Ar""bi:"bi:El]}, or, more easily, by pronouncing it similarly to gerbil \textipa{["\t{dZ}\textrhookschwa:b@l]}.} to indicate a Gamma-Ray Burst from a BLazar.\\
The paper is organized as follows. In Sect. \ref{LATdata}, I introduce the LAT data used in the work. In Sect. \ref{flareID}, I present how a GRBBL is identified in a light curve and how this process is automatized. In Sect. \ref{flarefit}, I present the fitting function used to parametrize GRBBLs. In Sect. \ref{flareclass}, I use unsupervised machine learning techniques to classify the GRBBLs. A discussion of the findings is provided in Sect. \ref{discussion}, before the concluding remarks in Sect. \ref{conclusions}.\\

\section{Fermi-LAT data}   
\label{LATdata}

The \textit{Fermi}-LAT light curves used in this project are retrieved from the LAT light curve repository \citep{LATLCR}. The download was done on August, 5, 2024. To downsize the amount of data to download, I preselected the AGNs of interest, by starting from the fourth catalog of AGNs detected by \textit{Fermi}-LAT \citep[4LAC;][]{4LAC}, and in particular its data release 3, limited to high Galactic latitude. From this source list, I filtered out all non-blazar AGNs (radio galaxies, narrow-line Seyfert-1 galaxies, Seyferts, compact-symmetric-sources, steep-spectrum radio-quasars, and AGNs without classification) and then all blazars with an unknown redshift, because a critical later step is the conversion from fluxes to luminosities. Initially, the redshift information was extracted directly from the 4LAC catalog. The LAT light curve repository was then looked at to see if the source was among the ones for which a light curve is provided. These steps define the data selection: included in this work are all the 4LAC blazars with a known redshift and whose light curve is available in the LAT light curve repository. This preliminary list of targets is composed of 846 blazars. I further cleaned the sample by double-checking the redshifts provided in the 4LAC catalog, manually inspecting the redshift references in the SIMBAD, NED, and ZBLLAC \citep{zbllac} databases. This step represents a major cleaning of the dataset: 114 blazars were removed due to uncertain redshift, about $13\%$ of the total (see Appendix \ref{app2}). For 22 of them, the redshift was modified. \\

\begin{figure}[t!]
    \centering
    \includegraphics[width=0.9\linewidth]{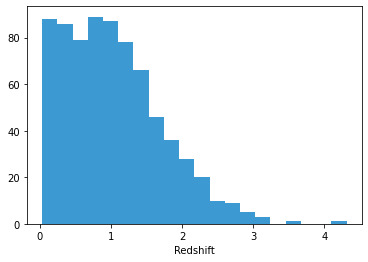}
    \caption{Redshift distribution of the blazars included in the work.}
    \label{fig:redshift}
\end{figure}

The final source list is composed of 732 blazars; it is provided in Appendix \ref{app1}. The redshift distribution is shown in Fig. \ref{fig:redshift}. Given that my final goal is to classify GRBBLs on the basis of their observational properties, I explicitly removed the information on the blazar class, flat-spectrum radio quasars (FSRQs) or BL Lacertae objects (BL Lacs). The reason is that this information is a label, and the classifier will obviously use it to class the events. On the other hand, it is interesting to see a posteriori how the blazar classes are related to the GRBBL classification. Hence, I only looked into blazar classes when I built my source list, and once the classification was done. Among the sources included in my final list, the large majority (527/732, 72$\%$) are FSRQs, followed by BL Lacs (168/732, 23$\%$), and blazars of unknown type (37/732, 5$\%$). From the LAT light curve repository, I downloaded three types of light curves, with different time binnings: three days; one week; and one month. The photon index was left free to vary during the light curve computation, and hence for every source and time binning I have two time series: the integral energy flux and the photon index versus time. The light curve repository provides flux points whenever the test-statistic is larger than a threshold value (chosen here to be 1); otherwise, an upper limit is provided. Before proceeding with the study, I corrected for the redshift: the time axis was divided by $(1+z)$, and the flux was converted to luminosity by multiplying it by $4\pi d_L^2$, where $d_L$ is the luminosity distance and was computed from the redshift assuming the cosmological parameters provided by \citet{Planck18}. It is important to highlight that here I am not correcting for the Doppler factor of the jet, because it is unknown for each individual object. Any dispersion in the distribution of the Doppler factors in the population will thus propagate down to the final result, and appear as a dispersion in the various quantities that I investigate.\\

Some caveats are provided in the LAT light curve repository paper and website. The most relevant one for this work is related to the presence of non-converging fits that result in outliers in the time series either in flux, or photon index, or both. These outliers were removed by performing rough cuts in the time series, and excluding data with photon indexes larger than $-1$ and smaller than $-5$, and fluxes larger than $10^{-7}$ TeV cm$^{-2}$ s$^{-1}$. Following the light curve repository website, I also investigated how the fluxes correlate with the test statistics (TS): I applied a second cut by removing all data points for which the flux over TS ratio is larger than 100 times the average of this ratio for the whole light curve. Holes in the light curves introduced by these cuts are not problematic for this study: I am interested in very high-quality light curves, which allow for the best possible characterization of a GRBBL; as is discussed later, only light curves with no gaps will be used to characterize GRBBLs, so this step simply ensures that outliers are removed from the data and not picked up by the flare identification step that is described in the next section.\\

\begin{figure}[t!]
    \centering
    \includegraphics[width=1.\linewidth]{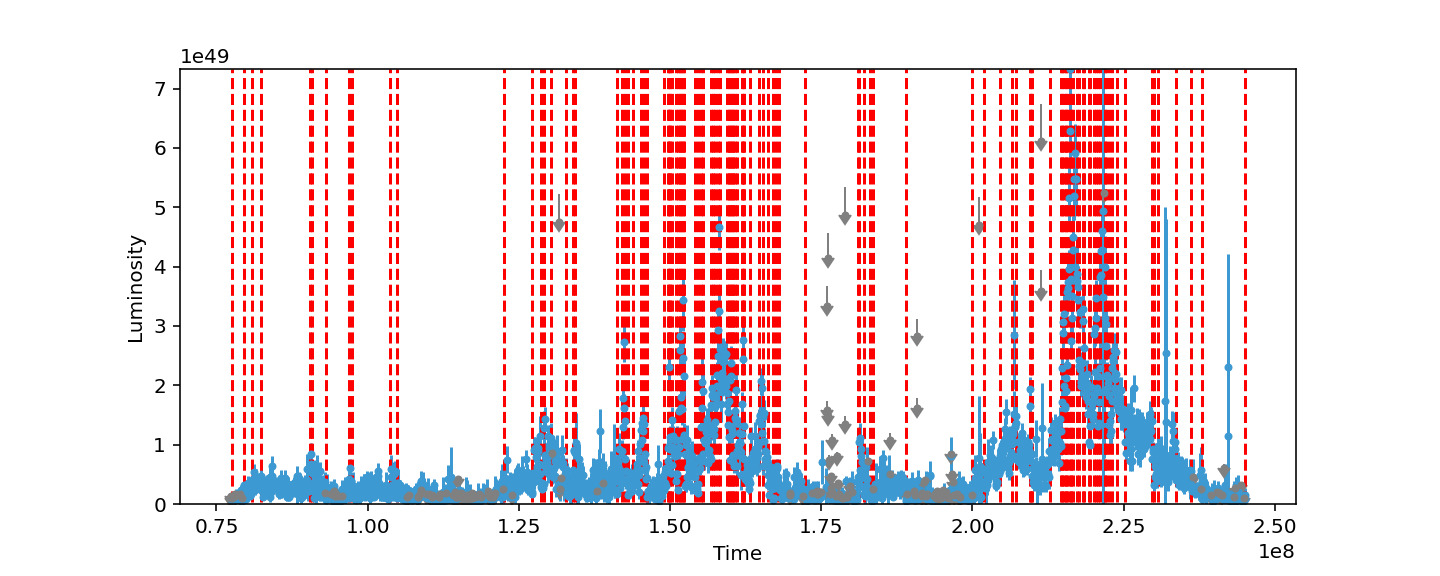}
    \includegraphics[width=1.\linewidth]{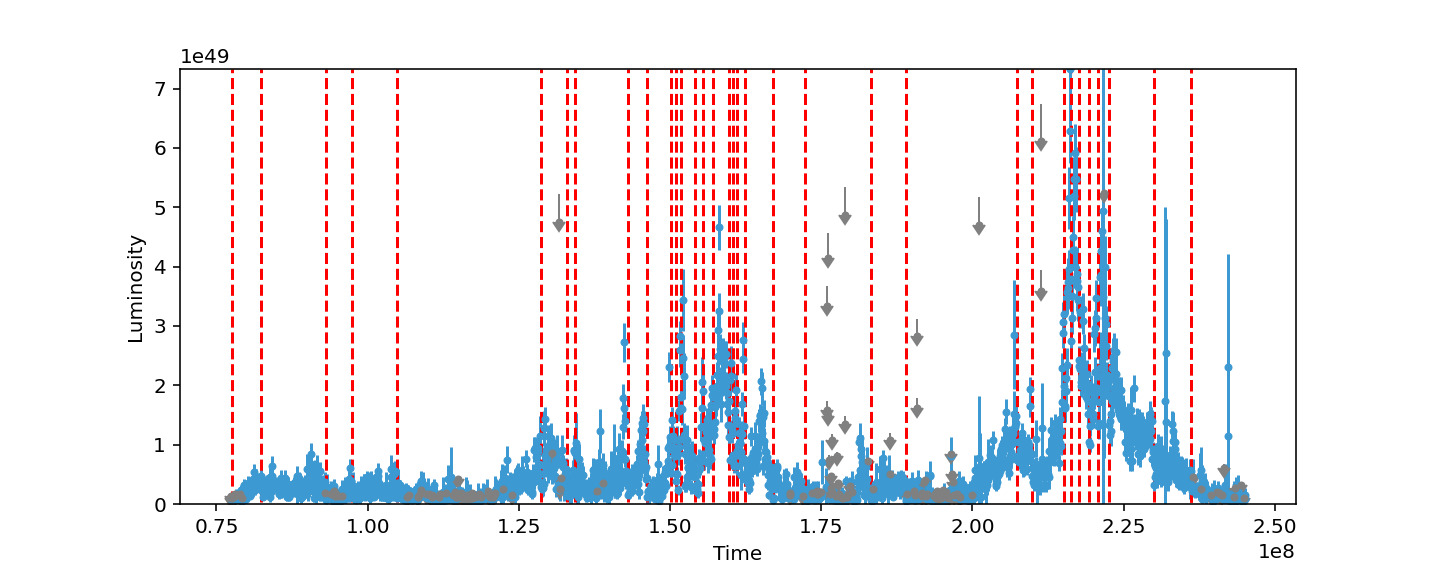}
    \caption{Light curve (luminosity vs time) of 4FGL~J0108.6+0134, with a three day binning. The vertical red lines represent the segmentation by the Bayesian blocks algorithm with $p_{BB} = 0.1$ (top), and by the merged superblocks (bottom, see text for their definition).}
    \label{fig:sample_LC}
\end{figure}

\section{GRBBL identification}
\label{flareID}

Blazar light curves are characterized by variability at all timescales, and even though a flare is often clearly visible, it is superposed on a varying continuum. In this sense, the task of identifying a GRBBL and applying a time cut to study its evolution is much harder than for catastrophic transients such as GRBs and SNs. 
A common tool to identify flux changes in time series is Bayesian blocks \citep{Scargle13}. The algorithm identifies significant changes in a light curve and provides the best segmentation into several blocks. I used the Astropy implementation of Bayesian blocks, with the “measures” option; that is, taking as an input the luminosity evolution and its uncertainty. The algorithm needs as an input a predefined false positive probability, $p_{BB}$, which impacts the number of blocks. This is effectively a free parameter of the algorithm, and I tested three values: $0.1$, $0.01$, and $0.001$. The first one was the one used in the fiducial test (because it results in a larger number of identified GRBBLs); the other two values were used to test how the conclusions of this work depend on this free parameter. In the top panel of Fig. \ref{fig:sample_LC}, I show the light curve of the blazar 4FGL~J0108.6+0134, and its segmentation in Bayesian blocks.\\

This first step only allows me to identify periods of time in which the luminosity is compatible with a constant, but I am still far from identifying GRBBLs. In the case, for example, of a well-sampled light curve that shows a simple increase and decrease in luminosity, with every bin computed from a high significance detection, the Bayesian block will identify every individual bin as a distinct state. I adopted the following procedure to identify flaring periods. Starting from the first block, I checked if the average of the next block luminosity was lower or higher. If higher, it meant that the peak of a potential GRBBL was yet to come, and I merged the blocks. I then proceeded to the next block, and, again, if the luminosity kept increasing, I merged the blocks. As soon as I detected a lower luminosity, I considered that I had passed a peak, and I then started looking into whether the luminosity kept decreasing, continuing the merging. Once I detected a new luminosity increase, I stopped the merging, and I called this a superblock. To increase the baseline and ease the fitting (see the next section), when moving to the next superblock I went back by one of the original Bayesian blocks, effectively allowing an overlap among them. By construction, any superblock contains a luminosity maximum. It could contain more than one, depending on how the original blocks have been determined. In the bottom panel of Fig. \ref{fig:sample_LC}, I again show the light curve of the blazar 4FGL~J0108.6+0134, this time with the segmentation in superblocks. This procedure was repeated for all blazars in the sample, and for all the time binnings.\\

\section{GRBBL characterization}
\label{flarefit}

GRBBL light curves are often parametrized with a phenomenological function characterized by an exponential rise and decay of the luminosities. The profile can be asymmetric, and hence two different timescales are introduced, one for the rising part and one for the decay part ($\tau_r$ and $\tau_d$, respectively). If the exponential is given with base 2, the two timescales are the doubling and halving times. The equation reads 
\begin{equation}
L(t) = L_{base} + \frac{L_{peak}}{2^{-\frac{t-t_0}{\tau_r}} + 2^{+\frac{t-t_0}{\tau_d}}}
\end{equation}
There are five free parameters: the two timescales; $t_0$ and $L_{peak}$, which represent, respectively, the time and luminosity where the two exponential functions meet; and  $L_{base}$, which is just a constant luminosity to fit the periods before and after the GRBBL. For a symmetric profile, $t_0$ represents the time of the maximum, and $L_{peak}$ is equal to twice the maximum luminosity. For an asymmetric profile, both values diverge from this simple expectation. \\

\begin{figure}[t!]
        \centering
        \begin{subfigure}{}
            \includegraphics[width=0.4\textwidth]{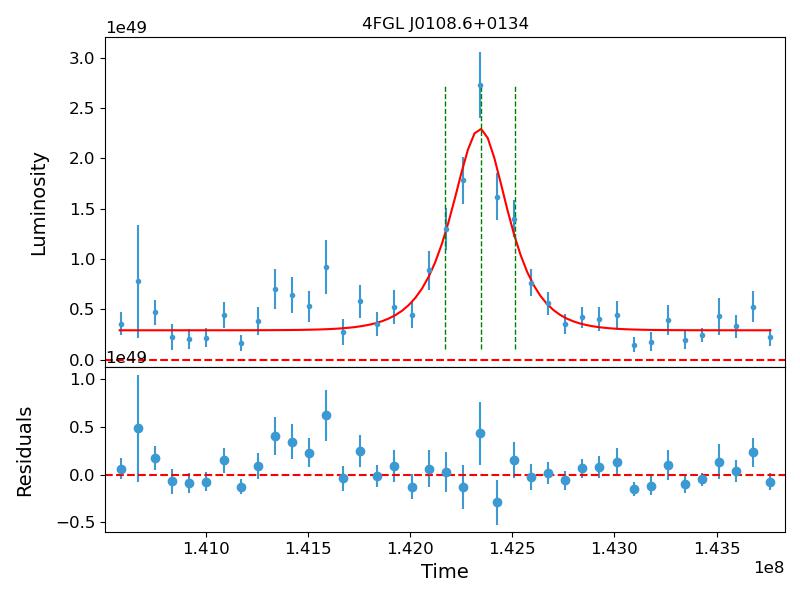}
        \end{subfigure}

        \begin{subfigure}{}
            \includegraphics[width=0.4\textwidth]{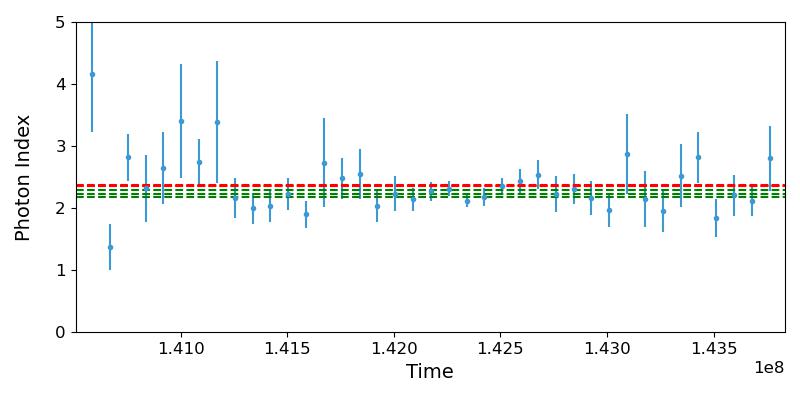}
        \end{subfigure}
      
        \begin{subfigure}{}
            \includegraphics[width=0.4\textwidth]{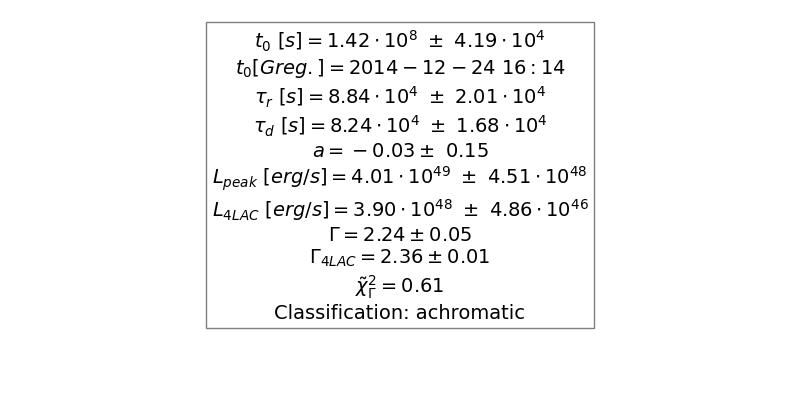}
        \end{subfigure}
        \caption{From top to bottom: Luminosity vs time (the vertical green lines indicate $t_0$ (the central one), $t_0 - 2\tau_r$ (on the left), and $t_0 + 2\tau_d$ (on the right); the red line represents the fit function); Residuals vs. time; Photon index vs. time (the horizontal green lines show the best-fit value of $\Gamma$ plus or minus its uncertainty; the horizontal red lines show the same but for $\Gamma_{4LAC}$); and the parameter values (see text) for one of the GRBBLs identified in 4FGL~J0108.6+0134.}
        \label{fig:example_flare}
    \end{figure}

This function was automatically fit to each of the superblocks that contained at least five data points. The fitting algorithm used was \textit{curve\_fit}, part of the SciPy library, and the procedure was completely automatized, with no human check: the starting values for the five free parameters ($L_{base}$, $t_0$, $L_{peak}$, $\tau_r$, and $\tau_d$) were the luminosity average of the superblock, the time of the maximum, the luminosity of the maximum, $10^5$, and $10^5$, and I set a maximum number of iterations equal to $10^6$. If the fit did not converge within this number of iterations, I considered it to be unsuccessful and I skipped this superblock. In the top panel of Fig. \ref{fig:example_flare}, I show as an example one of the GRBBLs identified in the light curve of 4FGL~J0108.6+0134. The residuals shown in the second subplot indicate that the fit was successful. \\

\begin{figure}[t!]
    \centering
    \includegraphics[width=1\linewidth]{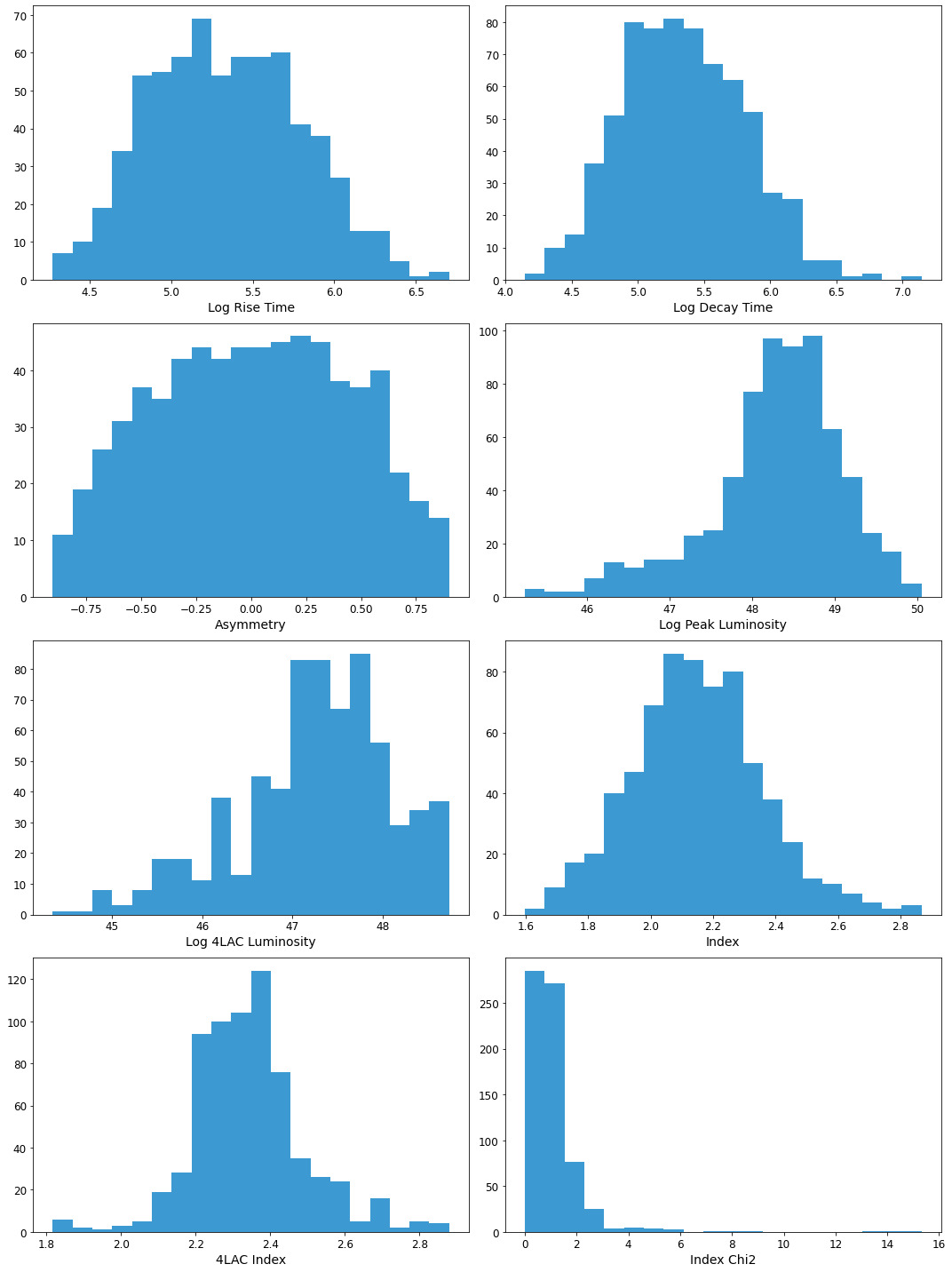}
    \caption{Histograms showing the distribution of the parameters passed to the classification algorithm.}
    \label{fig:histograms}
\end{figure}

\begin{figure*}[t!]
    \centering
    \includegraphics[width=1\linewidth]{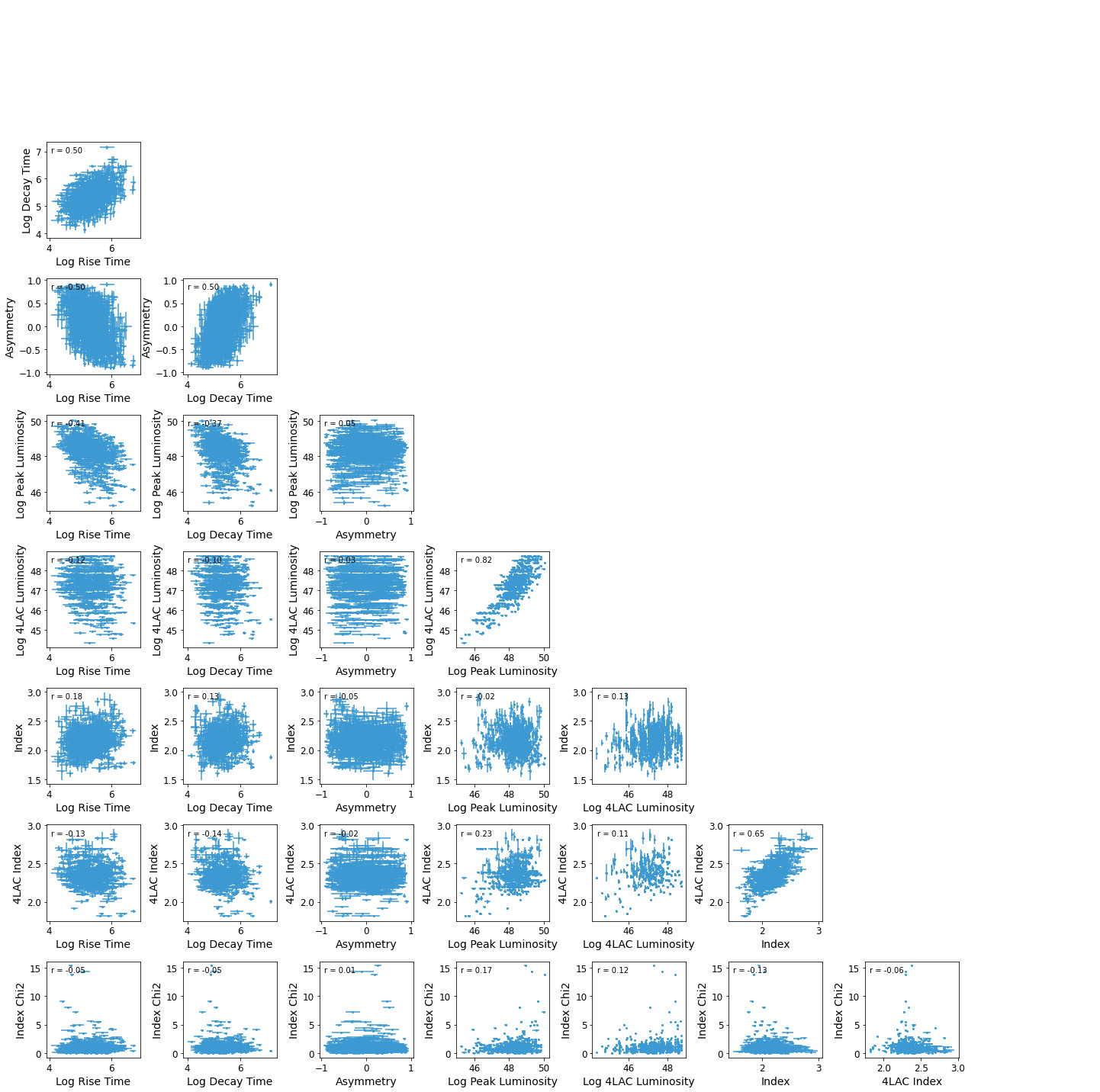}
    \caption{Corner plot showing all correlations among the eight parameters passed to the classification algorithm. }
    \label{fig:correlation}
\end{figure*}

In addition to the luminosity evolution over time, I am also interested in studying the evolution of the photon index that carries information about the energy distribution within the LAT energy band (the equivalent of color evolution in optical astronomy). The photon index does not vary as much as the luminosity, and I cannot fit the same function. To study spectral variability, I adopted the following procedure: I first chose an effective time interval, $t_{eff}$, which covers the duration of the GRBBL, defined between $t_0 - X\tau_r$ and $t_0 + X\tau_d$ (in the fiducial test, $X = 2$; I also tested $X = 1$ and $X = 3$); I then selected all time bins that fall within this time interval, adding also the last one before and the first one after; and then I fit  the photon index with a constant function. This procedure gave me two values: $\Gamma$, the average photon index during the GRBBL; as well as $\tilde{\chi^2_\Gamma}$, which is used to quantify if the photon index is compatible with a constant or not. In the central panel of Fig. \ref{fig:example_flare}, I show the evolution of the photon index for the same GRBBL of 4FGL~J0108.6+0134, and its fit constant value. In this case, it can be seen that there is no spectral variability. \\

Once all fits had been performed, I filtered them to make sure that only high-quality results passed this step and could be used for the classification. As a first filter, I removed all results for which there was at least a missing data point within the $t_{eff}$. Here, the most likely reason is that the missing point corresponds to an upper limit that I do not use in the fitting procedure, adding an unknown bias to the fit results. I also made sure that $t_{eff}$ was comprised within the superblock. As an additional filter, I removed all results for which $L_{peak}$ was not statistically different from $L_{base}$. Here, I used an arbitrary threshold value of $X\sigma$ ($X$ was chosen to be the same one that enters into the definition of $t_{eff}$, and for the fiducial test is equal to 2): only GRBBLs that fulfill $L_{peak} - X \sigma_{L_{peak}} > L_{base} + X \sigma_{L_{base}}$ pass this cut. I then filtered out all GRBBLs for which $L_{peak}$, $\tau_r$, or $\tau_d$ was consistent with zero at $X\sigma$. Lastly, I computed the residuals of the data to the fitting function, and I accepted only GRBBLs that have residuals within $t_{eff}$ at less than $3 \sigma$.\\

This procedure allowed me to automatically fit all superblocks and select only GRBBLs with high-quality light curves.  Given that I was working with light curves in three different binnings, many of the GRBBLs were identified more than once. For all GRBBLs that had more than one characterization, I selected the fit that had the smallest relative errors on the parameters. For the fiducial test ($p_{BB} = 0.1$ and $X = 2$), I ended up with 679 GRBBLs (see Appendix \ref{app3}). I passed to the classifier (see the next section) eight variables: $L_{peak}$, $\tau_r$, and $\tau_d$ from the light curve fit (I excluded $t_0$, which is not relevant, and $L_{base}$, which is only used as a baseline during the fit); $\Gamma$ and $\tilde{\chi^2_\Gamma}$ from the photon index fit; the average values of luminosity and photon index from the 4LAC catalog, $L_{4LAC}$, and $\Gamma_{4LAC}$; and lastly, an asymmetry parameter that I explicitly computed, 
\begin{equation}
a = \frac{\tau_d - \tau_r}{\tau_d + \tau_r}
.\end{equation}
The distributions of these eight parameters are shown in Fig. \ref{fig:histograms}. In Fig. \ref{fig:correlation}, I show all of the correlations among these parameters in the fiducial dataset.\\

\section{GRBBL classification}
\label{flareclass}

\begin{figure}[ht!]
        \centering
        \begin{subfigure}{}
            \includegraphics[width=0.4\textwidth]{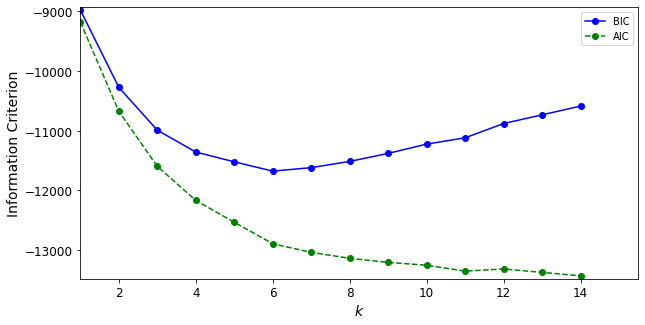}
        \end{subfigure}
        \begin{subfigure}{}
            \includegraphics[width=0.4\textwidth]{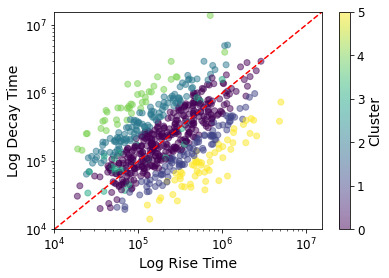}
        \end{subfigure}
               
        \begin{subfigure}{}
            \includegraphics[width=0.4\textwidth]{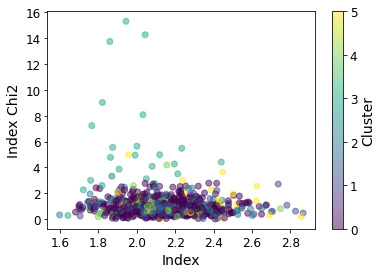}
        \end{subfigure}
        \caption{Results from the Gaussian mixture algorithm. From top to bottom: BIC (blue) and AIC (green) curves as a function of the number of clusters $k$; correlation plot for $\tau_d$ vs $\tau_r$ (same as in Fig. \ref{fig:correlation}, error bars have been removed for clarity) color-coded by the most likely cluster for $k=6$; same as before, but for the $\Gamma$ vs $\tilde{\chi^2_\Gamma}$ correlation.}
        \label{fig:GM}
    \end{figure}

The final goal of this study is to see if there are any similarities or differences within the population of GRBBLs. I approached this problem by looking at clusters in the eight-dimensional parameter space obtained in the previous section. Visual inspection of the correlation plots in Fig. \ref{fig:correlation} indicates that the dataset is rather homogeneous, with no clear clustering, but I would like to quantify this statement. The task of identifying clusters in a dataset is a well-known problem in unsupervised machine learning, and several algorithms exist. One of the most common tools is Gaussian mixtures, which identifies clusters in the dataset with the only explicit assumption being that their distribution is Gaussian. This algorithm does not take into account uncertainties in the variables, which makes it unsuitable for this study. Nonetheless, given that this is the first time a study of the global properties of GRBBLs has been performed, I consider it instructive for the reader to see the result of the application of Gaussian mixtures to the GRBBLs that I have identified and characterized. I used the scikit-learn implementation of Gaussian mixtures, and I passed as inputs the eight variables described above. Given that some of the parameters have very large values, and their distribution is very skewed on a linear scale, I passed $\tau_r$ and $\tau_d$, and $L_{peak}$ and $L_{4LAC}$, as their logarithms with base 10. All variables were then normalized with MinMaxScaler before computing Gaussian mixtures. The algorithm does not calculate the optimal number of clusters, $k$; rather, it computes the best classification for a given value of $k$. I tested $k$ from 1 to 14, and for each value I computed both the Bayesian information criterion (BIC) and the Akaike information criterion (AIC). The optimal value of  $k$ is then the one that minimizes the information criteria. The results are shown in Fig. \ref{fig:GM}. As can be seen, the BIC and AIC do not agree: the first one goes through a minimum at $k=6$, while the second one keeps improving (i.e., overfitting) and finds a minimum at $k=14$. The middle and bottom panels of Fig. \ref{fig:GM} show two of the correlation plots, color-coded by the most likely cluster for the optimal solution according to the BIC. The algorithm identifies five clusters according to their asymmetry, and then a sixth cluster is identified as the GRBBLs with high $\tilde{\chi^2_\Gamma}$.\\

\begin{figure}[t!]
        \centering
        \begin{subfigure}{}
            \includegraphics[width=0.4\textwidth]{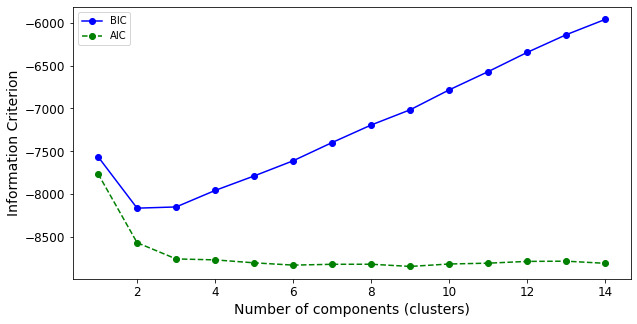}
        \end{subfigure}
        \begin{subfigure}{}
            \includegraphics[width=0.4\textwidth]{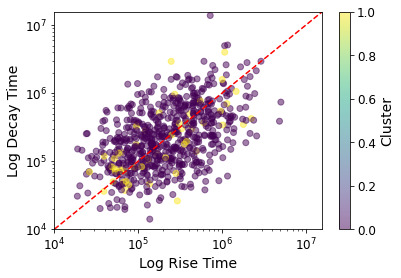}
        \end{subfigure}
               
        \begin{subfigure}{}
            \includegraphics[width=0.4\textwidth]{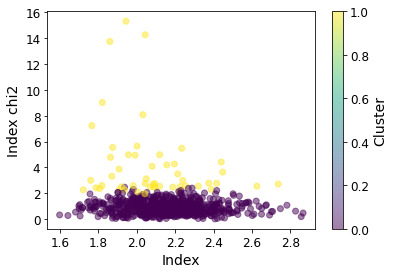}
        \end{subfigure}
        \caption{Same as in Fig. \ref{fig:GM} but for the extreme deconvolution algorithm.}
        \label{fig:XD}
    \end{figure}

\begin{figure}[t!]
        \centering
        \begin{subfigure}{}
            \includegraphics[width=0.4\textwidth]{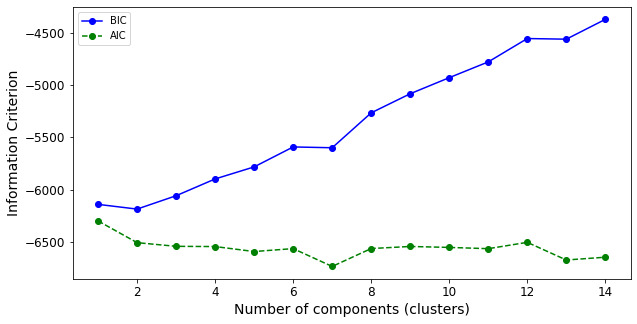}
        \end{subfigure}
        \begin{subfigure}{}
            \includegraphics[width=0.4\textwidth]{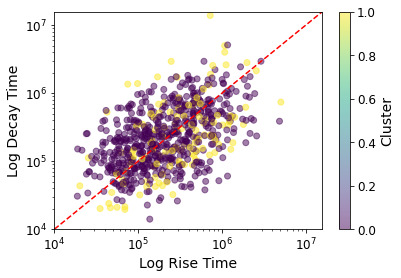}
        \end{subfigure}
               
        \begin{subfigure}{}
            \includegraphics[width=0.4\textwidth]{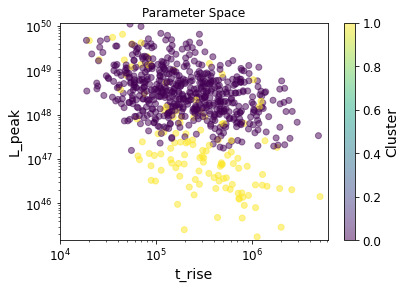}
        \end{subfigure}
        \caption{Same as in Fig. \ref{fig:XD} but for the extreme deconvolution algorithm run on only seven variables, excluding $\tilde{\chi^2_\Gamma}$.}
        \label{fig:XD_2}
    \end{figure}

Gaussian mixtures are shown here as an instructive first look into the classification, but the fact that the uncertainties on the model parameters are not taken into account makes any conclusion unreliable. The next step is to use extreme deconvolution \citep{Bovy}, which also takes the uncertainty on the variables as an input. A word of caution is needed here in the way the uncertainties are computed. The algorithm accepts only symmetric uncertainties, while when transforming from linear to logarithmic variables the uncertainties become asymmetric. Passing linear variables is clearly not a solution for the times and luminosities, which cover several orders of magnitude. I thus kept working with the logarithms, and passed as the uncertainty the average of the positive and negative uncertainties. The result of the extreme deconvolution classification is shown in Fig. \ref{fig:XD}. In this case as well, BIC goes through a clear minimum, while AIC tries to maximize $k$, indicating that it is not a good tool to identify the optimal number of clusters. For all the following tests in this work, I only consider BIC as the metric to estimate $k$. The algorithm identifies only two clusters, and the discriminant value is $\tilde{\chi^2_\Gamma}$. In this first test, GRBBLs cannot be classified as a function of their timescales, luminosities, or photon indexes, and the only classification is in chromatic and achromatic GRBBLs. To quantify how well separated the two clusters are, I checked several metrics widely used to study clustering of a dataset. The first one is the silhouette score, which estimates how each element of a cluster is similar to the other elements of its class (a strong clustering has a silhouette score close to 1); the second one is the Davies-Bouldin index (DB index; a strong cluster has a low DB value); and the third one the Calinski–Harabasz index (CH index; a strong cluster has a large CH value). For the fiducial dataset, the classification into chromatic and achromatic GRBBLs has a silhouette score of 0.11, a DB index of 4.09, and a CH index of 10.99: all metrics indicate that the separation between clusters is rather weak, with a significant overlap between the two classes.  As a second test, extreme deconvolution was run without $\tilde{\chi^2_\Gamma}$, to see if other weaker clusters were detected once the chromaticity was removed. The result of the extreme deconvolution classification is shown in Fig. \ref{fig:XD_2}. The algorithm again identifies two clusters, with a preference for $k=2$ over $k=1$ by $\Delta (BIC) = 45$ (which is significant, although admittedly lower than for the classification in chromatic versus achromatic. that had $\Delta (BIC) = 599$). This time, the parameter driving the classification seems to be $L_{peak}$ (see bottom plot in Fig. \ref{fig:XD_2}): the first cluster (in the following, type-1) has a high luminosity with a small dispersion, while the second one (type-2) spans the whole range in luminosities and explain the events at low $L_{peak}$. This can also be seen in the histograms in Fig. \ref{fig:histograms}: the one for $L_{peak}$ indeed shows a large queue at low values that the classifier attributes to the second class. Here, the metrics again indicate weakly separated clusters, although better defined than before: this classification of GRBBLs has a silhouette score of 0.24, a DB index of 2.33, and a CH index of 79.66. These results are of course only valid for the fiducial dataset, which has been produced for two specific values of the two parameters, $p_{BB}$ and $X$. I ran the classifier as a function of these parameters, and the results are provided in Table \ref{table:furthertests}. As can be seen, the classification as chromatic and achromatic GRBBLs is solid as a function of $p_{BB}$ and $X$, while the classification as type-1 and type-2 GRBBLs disappears for larger values of $X$ and smaller values of $p_{BB}$.\\

\section{Discussion} 
\label{discussion}

The main result of this work is that the population of GRBBLs is rather homogeneous: a variety of well-separated classes of GRBBLs does not emerge, and nor do clear outliers. When ignoring the photon index evolution, there are at most two classes of GRBBLs, but this classification is not seen in all datasets, and depends on the way the list of GRBBLs is compiled and filtered (i.e., the parameters $p_{BB}$ and $X$). This is a major piece of information for theoretical models aiming to reproduce \gray\ emission from blazars: they should be able to fit not just one of the light curves, but virtually all the light curves, with a continuous variation in the model parameters. Models should also be able to explain the existence of chromatic GRBBLs and their relative occurrence within the overall population. In the fiducial test, chromatic GRBBLs number only 51 out of 679, indicating that they represent rare occurrences within the whole population (see Appendix \ref{app4}). On this point, it is important to underline that I have not included the information on the TS of the detection in the study. It is likely that the GRBBLs that are more clearly classified as chromatic are the ones that have the best estimates of the photon index evolution. Or, stated in a different way, other GRBBLs could be intrinsically chromatic, but due to the relatively lower significance of their detection their photon index evolution is compatible with a constant. This is also suggested by the clustering metrics, which all indicate a significant overlap between the two clusters and suggest a smooth transition from one to the other. By looking at the distribution of $\tilde{\chi^2_\Gamma}$ (see Fig. \ref{fig:histograms}), it is clear that it behaves differently from the other variables, so it is not surprising that the extreme deconvolution uses this information to discriminate among GRBBLs. It can also be interpreted in the following way: the classifier recognizes the GRBBLs with a high value of $\tilde{\chi^2_\Gamma}$ as outliers in an otherwise homogeneous population.\\ 

As was discussed in the introduction, I did not include the information on blazar classes in the classifier, but it is interesting to measure statistics about the blazar classes a posteriori. I first investigated whether the GRBBLs are more likely found in FSRQs or BL Lacs (all calculations were computed only in the fiducial dataset). The 679 GRBBLs are found in 236 blazars. Of these, 201 (85$\%$) are FSRQs, 29 are BL Lacs (12$\%$), and 6 are blazars of an unknown type (2.5$\%$). With respect to the initial source list, there is a significant increase in FSRQs (from 72 to 85$\%$), indicating that GRBBLs are more likely found in this blazar class. It is interesting to see if the classifications have any overlap with the blazar classes. The 51 chromatic GRBBLs are found in 37 blazars. Of these, 31 (84$\%$) are FSRQs, 5 are BL Lacs (14$\%$), and 1 is a blazar of unknown type (3$\%$). These percentages are remarkably similar to the ones for the whole population of GRBBLs, indicating that chromatic GRBBLs are not biased toward any of the blazar classes. On the other hand, a closer look at the chromatic events might suggest that they are related to VHE detections  by Cherenkov telescopes: 50$\%$ (5/10) of the VHE FSRQs are included in the chromatic list, and out of the 5 BL Lacs in the list, 3 (60$\%$) are known VHE emitters. Concerning the classification into type-1 and type-2 GRBBLs, the 122 type-2 GRBBLs are found in 63 blazars. Of these, 40 (63$\%$) are FSRQs, 20 are BL Lacs (32$\%$), and 3 are blazars of unknown type (5$\%$). In this case, there is a clear increase in the BL Lacs, indicating that this class of objects is more represented among the type-2 class. This finding is in line with the fact that the main parameter that enters into the type-1/2 classification is the luminosity, which is known to be larger in FSRQs.\\
I ran the same algorithm for two parameters, $p_{BB}$ and $X$. The first one represents the false alert probability when computing the first segmentation of the light curve. A stricter (lower) value of $p_{BB}$ implies a smaller number of final blocks, and thus a smaller number of identified GRBBLs. It is a rather significant effect: by reducing $p_{BB}$ by a factor of 10, I  miss 19$\%$ of the GRBBLs. These lost events are not of bad quality (they pass the quality checks of the fit) and represent a true loss. What happens is that some of the blocks I work with end up containing more than one maximum, but the algorithm is written to search for one maximum per superblock. For this reason, being too strict on $p_{BB}$ is not particularly useful, and this explains the choice of 10$\%$ as the fiducial one. The second parameter, $X$, is used to define the duration of the GRBBL ($t_{eff}$) as well as to select high-quality fits (by cutting at $X\sigma$ the fit parameters). The effect of this parameter is much larger than the previous one. If I am stricter ($X=3$), I cut 64$\%$ of the GRBBLs and, this being a cut on the quality of the fit, I end up with a much cleaner sample. Conversely, a looser cut allows for many more GRBBLs to be accepted, increasing the sample by 130$\%$. On the other hand, the looser cut lowers the quality of the dataset, and it is not surprising that the results from the classifier are modified: the extreme deconvolution run on all eight parameters identifies three clusters, corresponding to the chromatic, type-1, and type-2 GRBBLs; when run on seven parameters it again finds three clusters, corresponding to type-2 GRBBLs, and then splitting the sample rather arbitrarily into positive versus negative asymmetry. The choice of $X = 2$ as fiducial value thus represents an average between these two extremes.\\

Although there are not obvious outliers in the population, it is interesting to mention the record holders (in the fiducial test): 
\begin{itemize}
\item the fastest GRBBLs are: in 4FGL J2328.3-4036 on April 27, 2012 (with the fastest $\tau_r = 1.88 \cdot 10^4$ s), and in 4FGL J2311.0+3425 on January 4, 2022 (with the fastest $\tau_d = 1.41 \cdot 10^4$ s);
\item the slowest GRBBLs are: in 4FGL J0809.8+5218 on October 10, 2011 (with the slowest $\tau_r = 5.06 \cdot 10^6$ s), and in 4FGL J1231.7+2847 on May 11, 2009 (with the slowest $\tau_d = 1.39 \cdot 10^7$ s);
\item the most asymmetric GRBBLs are: in 4FGL J1427.9-4206 on March 10, 2017 (with the smallest $a = -0.90$ s), and in 4FGL J1231.7+2847 on May 11, 2009 (with the largest $a = +0.90$ s);
\item the most luminous GRBBL is in 4FGL J2232.6+1143 on February 11, 2017  (with the largest $L_{peak} = 1.11 \cdot 10^{50}$ erg/s);
\item the least luminous GRBBL is in 4FGL J1517.7-2422 on October 25, 2022 (with the smallest $L_{peak} = 1.76 \cdot 10^{45}$ erg/s);
\item the softest GRBBL is in 4FGL J1229.0+0202 on April 14, 2011  (with the largest $\Gamma = 2.87$ );
\item the hardest GRBBL is in 4FGL J0719.3+3307 on January 19, 2015 (with the smallest $\Gamma = 1.60$);
\item the highest contrast GRBBL (defined as the largest $L_{peak} / L_{4LAC}$) is in 4FGL J0112.8+3208 on January 3, 2023 (with the largest $L_{peak} / L_{4LAC} = 311$);
\item the lowest contrast GRBBL is in 4FGL J0210.7-5101 on January 11, 2011 (with the smallest $L_{peak} / L_{4LAC} = 0.57$);
\item the most chromatic GRBBL is in 4FGL J1512.8-0906 on August 10, 2015 (with the largest $\tilde{\chi^2_\Gamma} = 15.32$).
\end{itemize}

The correlations among the GRBBL parameters are shown in Fig. \ref{fig:correlation}, together with the Pearson coefficients. The majority of the variables are not correlated, with some notable exceptions. There is a positive correlation between $\tau_{r}$ and $\tau_{d}$, which indicates that GRBBLs are more likely symmetric, as can be seen from the histogram of the $a$ parameter. The asymmetry is also correlated with both $t_r$ and $t_d$. This comes directly from the definition of $a$ and the previous correlation: if a GRBBL has a fast $t_r$, at the rising queue of the distribution, it will be positively asymmetric, or at most symmetric; it is very unlikely that it will have an even faster $t_d$. The same consideration applies for the correlation between $a$ and $\tau_d$.
Each of the quantities extracted from the 4LAC catalog, correlates with its counterpart during the GRBBL: $L_{0}$ correlates with $L_{4LAC}$, and $\Gamma$ correlates with $\Gamma_{4LAC}$. In the first case, the peak luminosity is (as was expected) systematically above the equality line. The contrast ranges from 0.6 to 311 and can be used as a typical range for how much brighter GRBBLs are compared to the baseline flux. The correlation between the indexes is also particularly interesting: the index during the GRBBL follows the average index, but it is typically harder than the average. This is true not just for the achromatic GRBBLs, but for all of them. 
Finally, there is evidence (r=-0.37 and -0.41, respectively) for an anticorrelation between both $\tau_{d}$ and $\tau_{r}$, and $L_{peak}$: the brightest GRBBLs are typically the fastest. This is certainly the most interesting of the correlations that I identified, and it deserves to be discussed in separate works. It is obvious that, if there is indeed a link between a timescale and a luminosity, GRBBLs can then be used as cosmological probes. The plot shows that the dispersion is significant, and at first sight GRBBLs cannot compete with other probes such as SNs. But the fact that they can be detected up to larger redshifts makes this simple correlation worth deeper investigations. It will first be important to make sure that it is a genuine correlation and not a selection effect, and then study it further to see if the scatter can be reduced.\\

This work represents the first tentative attempt to study GRBBLs as a population, and there are certainly many ways in which it can be improved. The first limitation is the data access: here, I have only analyzed light curves available in the LAT light curve repository, but the same analysis should be done on all 4LAC sources. A critical aspect is also the availability of the redshift information: many GRBBLs have been removed because their distance estimate is not reliable. This work again highlights the importance of redshift campaigns for blazar studies. Staying on the data analysis, a critical aspect is the time binning of the time series. It is definitely interesting to investigate smaller time bins (one day, 12 hours) to see how the distribution of $\tau_r$ and $\tau_d$ behaves. Ideally, the best approach should be to adopt an adaptive binning, in order to enable a better evaluation of the flare, and to reduce biases coming from GRBLLs sampled at different TS levels. The flare characterization can also be improved, in particular by allowing for more complex fitting functions. The likely critical issue within the current work is that only very clear, isolated GRBBLs pass the filter. If several events overlap, fitting with a single rise and decay profile will probably fail to converge, or will result in high-significance residuals that then do not pass the quality cut. An important future direction could also be extending this work to non-blazar gamma-ray sources, to see how, for example, narrow-line Seyfert-1 galaxies or radio galaxies fit within this population. Lastly, this same approach should take the multiwavelength path. In the same way as for individual events we use multiwavelength information to improve our understanding of blazar physics, we should investigate the counterparts of GRBBLs at longer wavelengths and add dimensions to the parameter space to see if a classification emerges. While this approach is certainly difficult for non-survey instruments, it should be considered for optical surveys, or for sources that are regularly monitored by telescopes in radio or X-rays.

\begin{table}
\begin{center}
\begin{tabular}{c|c|c|c|c|c}
 & $p_{BB}$ & $X$ & $\#$GRBBLs & MinBIC\\
 \hline
 Fiducial & 0.1 & 2 & 679 & 2 \\
no $\tilde{\chi^2_\Gamma}$ & 0.1 & 2 & 679 & 2\\  
\hline
          & 0.1 & 1 & 1572    & 3 \\
no $\tilde{\chi^2_\Gamma}$   & 0.1 & 1 & 1572    & 3 \\      
\hline
          & 0.1 & 3 &  245    & 2 \\
no $\tilde{\chi^2_\Gamma}$          & 0.1 & 3 &  245    & 1 \\
\hline          
          & 0.01 & 2 & 547  & 2 \\
no $\tilde{\chi^2_\Gamma}$            & 0.01 & 2 & 547  & 2 \\
\hline
          & 0.001 & 2 & 445 & 2 \\ 
  no $\tilde{\chi^2_\Gamma}$         & 0.001 & 2 & 445 & 1 \\ 
 \hline
\end{tabular}
\caption{Results from the extreme deconvolution classification algorithm.}
\label{table:furthertests}
\end{center}
\end{table}   

\section{Conclusions}
\label{conclusions}

In this work, I have performed a comprehensive characterization and classification of all GRBBLs that I have been able to analyze and identify. It is the first study of this kind and it shows, as a proof of concept, that it is possible to perform a population study of GRBBLs, and gain an insight into the physics of blazars from their collective behavior. This type of study is common for blazars in their stationary state. Here, I show that this approach can be extended to light curves and to the properties of individual flares. The main conclusion is that the population of GRBBLs is rather homogeneous. A classification emerges between chromatic and achromatic GRBBLs. The chromatic ones are much rarer, show a clear evolution of the photon index during the flare, making it incompatible with a constant fit. This result is solid and does not depend on the parameters used to identify and characterize the events. If the information on the photon index evolution is removed, a second classification appears, mainly on the basis of the peak luminosity of the events. This classification seems less clear, primarily because it disappears when changing the parameters of the algorithm.\\

As a by-product of this project, I have identified for the first time a correlation between the rise and decay timescales and the peak luminosities of GRBBLs. Although admittedly the scatter is large, the correlation deserves to be further investigated, because it might pave the way for using GRBBLs as cosmological probes. \\ 

\section*{Data availability}
\label{table:sourcelist}
\label{table:removedlist}
\label{table:chromaticlist}

Tables A1, B1, and D1 are only available in electronic form at the CDS via anonymous ftp to \texttt{cdsarc.u-strasbg.fr} (130.79.128.5) or via 
\url{http://cdsweb.u-strasbg.fr/cgi-bin/qcat?J/A+A/}. Appendix \ref{app3} is available on Zenodo: \url{https://doi.org/10.5281/zenodo.15061676}.\\

\begin{acknowledgements} 
I thank Lea Heckmann and Enzo Oukacha for useful inputs on the classification algorithms, Janeth Valverde for feedback on the LAT light curve repository, and Paolo Goldoni for feedback on the redshift of the sources. I would also like to thank the anonymous referee, who provided insightful comments that significantly improved the work and the manuscript. This work used generative artificial intelligence (chat-GPT version 3.5 and 4.0) to speed-up the development of, optimize, and clean-up the python code. No generative artificial intelligence has been used to write the manuscript.
\end{acknowledgements}

\nocite{Pena21, Goldoni21, Archambault16, Shaw13, Sbarufatti05, Jones09, Shaw12, Titov13, Xu94, Garcia23, Dorigo22, Paiano20, cgrabs, Veron10, Paiano21}

\nocite{1WHSP, Shaw12, Fujinaga16, Pita14, Paiano17, Shaw13, Paiano17b, DAmmando24, Paiano16, Paiano20, Meisner10, 3HSP, Stadnik14, Jones09, Rau12, Landoni15, Goldoni21, Paiano17c, Nilsson18, Carswell74, Caccianiga, Richards09, Paggi14, Landoni13, Massaro14, Pena19, Sbarufatti05im, Desai19, Britzen07, Garcia23, cgrabs, Olmo22, Massaro13, Landoni14, Marleau07, Tinti06, Paiano19, Landoni13}

\bibliographystyle{aa} 
\bibliography{LAT_flares}

\appendix

\section{List of sources}
\label{app1}

In table A1 I provide the list of blazars analyzed in this work. \\

\section{List of sources removed due to uncertain redshift}
\label{app2}
In table B1 I provide the list of blazars removed due to uncertain redshift. In addition to the nominal redshift provided in the 4FGL, I indicate in the "Comment" column the reason why the source has been removed. \\

\section{GRBBL figures}
\label{app3}
In the following figures, I show the first ten GRBBLs in the fiducial dataset (ranked by RA and $t_0$). The complete list of 679 figures is available on Zenodo: \url{https://doi.org/10.5281/zenodo.15061676}.\\

% Loop over block IDs
\foreach \n in {1,...,10} {%
    \begin{figure}[ht!]
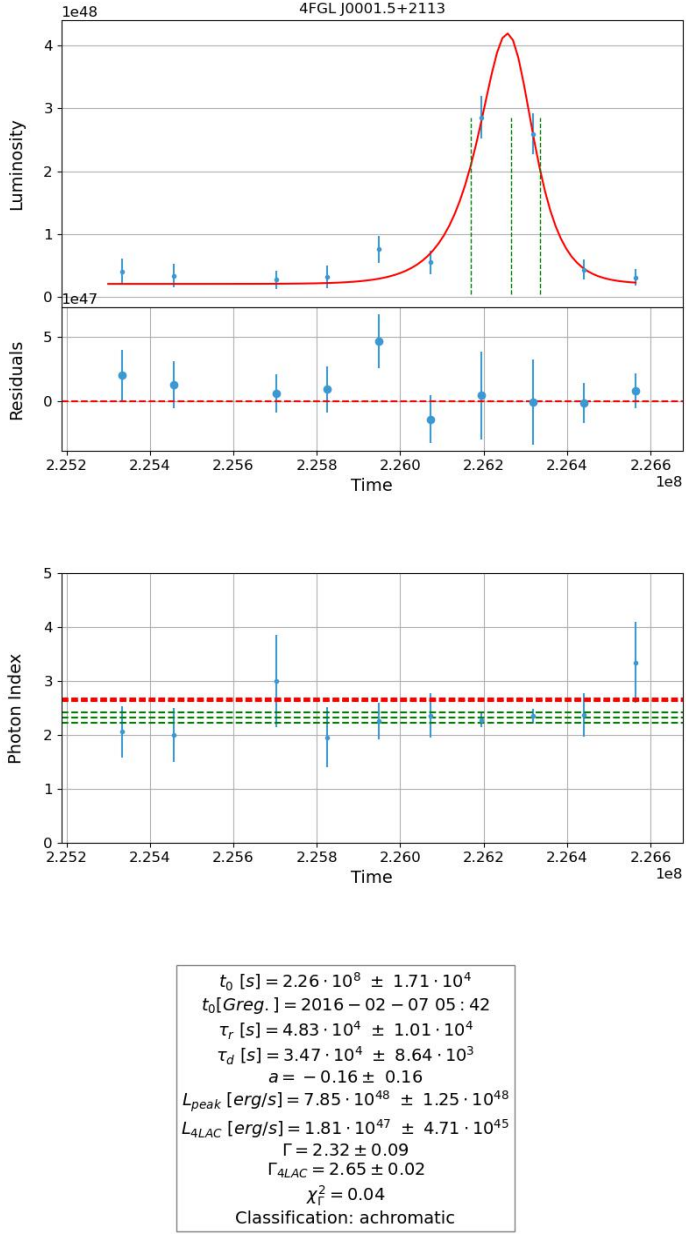
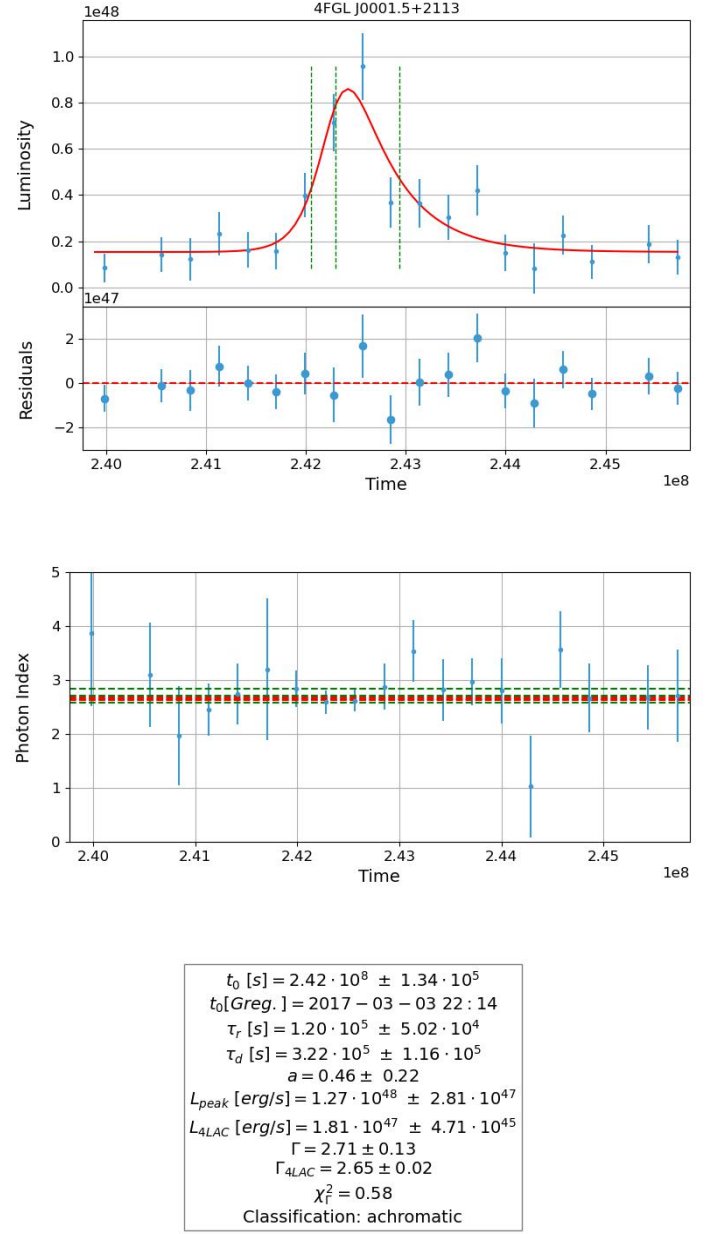

        \centering
        \begin{subfigure}{}
            \includegraphics[width=0.5\textwidth]{\n_luminosity.jpg}
        \end{subfigure}
        \hfill
        \begin{subfigure}{}
            \includegraphics[width=0.5\textwidth]{\n_photon_index.jpg}
        \end{subfigure}
        \hfill       
        \begin{subfigure}{}
            \includegraphics[width=0.5\textwidth]{\n_table.jpg}
        \end{subfigure}
        \caption{Same as Fig. \ref{fig:example_flare}, but for GRBBL number \n}
        \label{fig:flarefigures\n}
    \end{figure}
}

\section{Chromatic GRBBLs}
\label{app4}
In table D1 I provide the list of blazars that are classified as chromatic (in the fiducial dataset).
\end{document}